\documentstyle[preprint,aps,epsfig,floats]{revtex}

\newcommand{\be}{\begin{equation}}

\newcommand{\ee}{\end{equation}}

\newcommand{\bea}{\begin{eqnarray}}

\newcommand{\eea}{\end{eqnarray}}

\newcommand{\bml}{\begin{mathletters}}

\newcommand{\eml}{\end{mathletters}}

\begin{document}

\tighten

\preprint{DCPT-03/07}

\draft

%\twocolumn[\hsize\textwidth\columnwidth\hsize\csname @twocolumnfalse\endcsname

%%%%%%%%%%%%%%%%%%%%%%%%%%%%%%%%%%%%%%%%%%%%%%%%%%%%%%%%%%%%%%%%%%%%%%%%%%

%\wideabs{                       % Uncomment this line for two-column output

\title{Strings in de Sitter space  }

\renewcommand{\thefootnote}{\fnsymbol{footnote}}

\author{Eug\^enio R. Bezerra de Mello\footnote{emello@fisica.ufpb.br}}
\address{Departamento de F\'{\i}sica-CCEN, Universidade Federal da 
Para\'{\i}ba, 58.059-970, J. Pessoa, PB, C. Postal 5.008, Brazil}
\author{ Yves Brihaye\footnote{Yves.Brihaye@umh.ac.be}}
\address{Facult\'e des Sciences, Universit\'e de Mons-Hainaut,
 B-7000 Mons, Belgium}
\author{Betti Hartmann\footnote{Betti.Hartmann@durham.ac.uk}}
\address{Department of Mathematical Sciences, University
of Durham, Durham DH1 3LE, U.K.}
\date{\today}

\setlength{\footnotesep}{0.5\footnotesep}

\maketitle

%%%%%%%%%%%%%%%%%%%%%%%%%%%%%%%%%%%%%%%%%%%%%%%%%%%%%%%%%%%%%%%%%%%%%%%%%%

\begin{abstract}

We study both global as well as local (Nielsen-Olesen) strings
in de Sitter space. While these type of topological defects
have been studied in the background of a de Sitter metric previously, we study
here the full set of coupled equations.
We find only ``closed'' solutions. The behaviour of the metric tensor
of these solutions resembles that of ``supermassive'' strings with a curvature
singularity at the cosmological horizon.
For global strings (and the composite defect) we are able to construct 
solutions which are regular on the interval from the origin to
the cosmological horizon if the global string core lies completely inside the
horizon.
\end{abstract}


%%%%%%%%%%%%%%%%%%%%%%%%%%%%%%%%%%%%%%%%%%%%%%%%%%%%%%%%%%%%%%%%%%%%%%%%%%

%\newpage

\renewcommand{\thefootnote}{\arabic{footnote}}

\section{Introduction}
A number of different topological defects \cite{shell} are thought to have been 
formed during the phase transitions in the early universe. Depending on the topology
of the vacuum manifold ${\cal M}$ these are domain walls, strings, monopoles and textures corresponding
to the homotopy groups $\pi_0({\cal M})$, $\pi_1({\cal M})$, $\pi_2({\cal M})$ and $\pi_3({\cal M})$, respectively.
Cosmic strings \cite{shell,kibble} have always gained a lot of interest since they are thought
to be important for the structure formation in the universe due to their huge
energy per unit length (roughly $10^{21}\frac{kg}{m}$ for a string formed at
GUT scale $\approx 10^{16} GeV$).

A classical field theory model which has string-like solutions is the Abelian Higgs model \cite{no}. These solutions, also sometimes called ``vortices'', correspond to 
infinitely long objects. They have a core radius inverse proportional
to the Higgs boson mass and magnetic flux tubes with radius proportional to the 
inverse of the gauge boson mass.  Coupling the Abelian Higgs model minimally to gravity, the influence of the vortex on the geometry of space-time was investigated analytically \cite{Gar}. It was shown that far away from the core
of the string the space-time is Minkowski minus a wedge. It was 
also realised \cite{laguna} that if the vacuum expectation value of
the Higgs field is sufficiently large (corresponding
to strings having formed at a phase transition with energy scale much higher
than the GUT scale), then a different type of solution
is possible. These so-called ``supermassive strings'' exist only on a
finite interval of the radial coordinate and have a curvature singularity at
the maximal value of the radial coordinate. 
The existence of further solutions was investigated in a detailed numerical
analysis \cite{Christensen,yves}. It was found that the parameter space
is indeed divided by the curve of maximal angular deficit $2\pi$. 
If the deficit is smaller than $2\pi$, so-called ``open'', 
i.e. infinitely extended solutions were found. One is the above mentioned
cosmic string solution \cite{Gar} which however has a ``shadow'' solution
of Melvin-type for all values of the coupling constants. 
For deficit large than $2\pi$, only the ``supermassive'',
``closed'' solutions exist \cite{laguna}.

The static solutions of the model without gauge field, so-called global
strings, have also been studied \cite{cohen,harari,Gibbons,gregory1}. Like all global defects, the global string has a
long-range Goldstone field which leads to a divergent energy.
Moreover, the global string is characterised by a logarithmically
divergent deficit angle in contrast to the local
string which has a constant deficit angle. The coupling to gravity in the case
of the static global monopole leads to a singularity-free monopole solution
\cite{vilenkin} in the sense that while the energy is still
linearly divergent, the solid deficit angle is now finite. For
the static global string the corresponding singularity can not be removed 
by coupling the system
to gravity and only the assumption that the metric be time-dependent
removes the singularity \cite{gregory2}.

Since a number of astrophysical observations like e.g. the measurement of
redshifts of Type Ia supernovae \cite{super} has led scientists to believe that we live in a universe with positive cosmological constant, the study of topological
defects in de Sitter (dS) space seems interesting.
But it also is of interest from another point of view, namely the dS/CFT
correspondence \cite{strominger}. This correspondence 
suggests a holographic duality between gravity in a $d$-dimensional
dS space and a conformal field theory (CFT) ``living''
on the boundary of the dS spacetime and thus being $d-1$-dimensional.

Recently,  Nielsen-Olesen strings in the background 
of a $4$-dimensional de Sitter spacetime $dS_4$ have been studied \cite{GM}. 
However, to our knowledge, the {\it full} system of coupled matter and metric
field equations has not been studied yet. One of the aims of this
paper is the investigation of exactly this point.

Motivated by some recent work on
a composite system of a 
global and local monopole in curved space-time \cite{bbh,by,spi},
we investigate the composite system of a global and 
Nielsen-Olesen string in a curved space-time with cosmological constant
as well.

Our paper is organised as follows: we give the model and static, cylindrically
symmetric Ansatz in Section 2. We give the equations of motion in
Section 3. We discuss the pure Nielsen-Olesen solutions
in Section 4, the global string solutions
in Section 5 and the composite system of a global and Nielsen-Olesen
string in Section 6. We give our summary in Section 7.

\section{The Model}
%%%%%%%%%%%%%%%%%%%%%
The model which describes a gravitating Nielsen-Olesen string interacting with
a global one in the presence of a non-vanishing cosmological constant is 
given by the following action:

\begin{equation}
\label{action}
S=\int d^4 x \sqrt{-g} \left( \frac{1}{16\pi G} (R-2\Lambda) + {\cal L}_{NO}+ 
{\cal L}_{global} + {\cal L}_{inter} \right)
\end{equation}
where $R$ is the Ricci scalar, $G$ denotes Newton's constant and $\Lambda$ is the 
cosmological constant. The Lagrangian of the  Abelian Higgs model is given by 
\cite{no}:
\begin{equation}
{\cal L}_{NO}=\frac{1}{2}D_{\mu} \phi (D^{\mu} \phi)^*-\frac{1}{4} F_{\mu\nu} F^{\mu\nu}
-\frac{\lambda_1}{4}\left(\phi\phi^*-\eta^2_1\right)^2
\end{equation} 
with the covariant derivative $D_\mu=\nabla_{\mu}-ieA_{\mu}$ and the
field strength $F_{\mu\nu}=\partial_\mu A_\nu-\partial_\nu A_\mu$  of the U(1) gauge potential $A_{\mu}$ with coupling constant $e$.
$\phi$ is a complex scalar field (the Higgs field) with vacuum expectation value $\eta_1$ and self-coupling constant $\lambda_1$.
The Lagrangian of the global string reads \cite{shell}:
\begin{equation}
{\cal L}_{global}=\frac{1}{2}\partial_{\mu} \chi \partial^{\mu} \chi^* - \frac{\lambda_2}{4}\left(\chi\chi^*-\eta^2_2\right)^2
\end{equation} 
where $\chi$ is a complex scalar field (the Goldstone field) with vacuum expectation value $\eta_2$ and self-coupling $\lambda_2$.
Finally, following \cite{bbh} we introduce an extra potential which couples (with coupling constant $\lambda_3$)
the two sectors of the model directly to each other:
\begin{equation}
{\cal L}_{inter}=-\frac{\lambda_3}{4}\left(\phi\phi^*-\eta^2_1\right)\left(\chi\chi^*-\eta^2_2\right)
\end{equation}
Without this term, the global and local string would be coupled only indirectly 
over gravity. In this paper we will use units which $\hbar=c=1.$

\subsection{The Ansatz}

In the following we shall analyse the classical equations of motion associated
with the above system. In order to do that,
let us write down the matter and gravitational fields as shown below.
The most general, cylindrically symmetric line element invariant under boosts
along the $z-$direction is:
\begin{equation}
ds^2=N^2(\rho)dt^2-d\rho^2-L^2(\rho)d\varphi^2-N^2(\rho)dz^2 \ .
\end{equation}
The non-vanishing components of the Einstein tensor $G_{\mu\nu}$ then read:
\begin{eqnarray}
& & G_{tt}=-G_{zz}=\frac{N}{L}\left(L\partial_{\rho\rho}N + \partial_{\rho}N \partial_{\rho}L+
N\partial_{\rho\rho}L \right), \nonumber \\
& & G_{rr}=\frac{\partial_{\rho}N}{N^2L}\left(2\partial_{\rho}L N+\partial_{\rho}N L\right)\ \ ,
\ \ G_{\varphi\varphi}=\frac{L^2}{N^2}\left(2N\partial_{\rho\rho}N+(\partial_{\rho} N)^2\right) \ ,
\end{eqnarray}
where $\partial_{\rho}$ denotes the derivative with respect to $\rho$. 

For the matter and gauge fields, we have:
\begin{equation}
\phi(\rho,\varphi)=\eta_1 h(\rho)e^{i n\varphi} \ ,
\end{equation}
\begin{equation}
\chi(\rho,\varphi)=\eta_1 f(\rho)e^{i m\varphi} \ ,
\end{equation}
\begin{equation}
A_{\mu}dx^{\mu}=\frac {1}{e}(n-P(\rho)) d\varphi \ .
\end{equation}
$n$ and $m$ are integers indexing the vorticity of the Higgs and Goldstone fields, respectively,  around the $z-$axis.

Substituting the above configurations into the matter Lagrangian density 
${\cal L}_{M}={\cal L}_{NO}+ {\cal L}_{global} + {\cal L}_{inter}$, we obtain:
\begin{eqnarray}
{\cal L}_M&=&-\frac{\eta_1^2}2(\partial_{\rho}h(\rho))^2-
\frac{\eta_1^2}2(\partial_{\rho}f(\rho))^2-\frac{n^2}
{2e^2L^2(\rho)}(\partial_{\rho}P(\rho))^2\nonumber\\
&-&\frac{\eta_1^2n^2}{2L^2(\rho)}h^2(\rho)P^2(\rho)-
\frac{\eta_1^2m^2}{2L^2(\rho)}f^2(\rho)-
\frac{\lambda_1\eta_1^4}4(h^2(\rho)-1)^2\nonumber\\
&-&\frac{\lambda_2\eta_1^4}4(f^2(\rho)-q^2)^2
-\frac{\lambda_3\eta_1^4}2 (h^2(\rho)-1)(f^2(\rho)-q^2) \  .
\end{eqnarray}

\section{Equations of Motion}

We define the following dimensionless variable and function:
\begin{equation}
x=\sqrt{\lambda_1}\eta_1 \rho \ \ \ , \ \ \ L(x)=L(\rho)\eta_1\sqrt{\lambda_1} \ .
\end{equation}
Then, the total Lagrangian only depends on the following dimensionless coupling constants
\begin{equation}
\gamma=8\pi G\eta_1^2 \ \ , \ \alpha=e^2/\lambda_1 \ \ , \ \ 
q=\frac{\eta_2}{\eta_1} \ \ , \ \ {\bar\Lambda}=
\frac\Lambda{\lambda_1\eta_1^2} \ \ , \ \
\beta_i^2=\frac{\lambda_i}{\lambda_1} \ \ , \ \ i=1,2,3 \ .
\end{equation}

Varying (\ref{action}) with respect to the matter fields and metric functions, we obtain a system of five non-linear differential equations. The Euler-Lagrange equations for the matter field functions read:
\begin{equation}
\frac{(N^2Lh')'}{N^2L}=\frac{n^2}{L^2} hP^2+h(h^2-1)+\beta_3^2h(f^2-q^2) \ ,
\end{equation}
\begin{equation}
\frac{(N^2Lf')'}{N^2L}=\frac{m^2f}{L^2}+\beta_2^2f(f^2-q^2)+\beta_3^2
f(h^2-1) \  ,
\label{eqf}
\end{equation}
\begin{equation}
\frac{L}{N^2}\left(\frac{N^2P'}{L}\right)'=\alpha h^2P \ ,
\end{equation}
while the Einstein equations
\begin{equation}
G_{\mu\nu}+\bar\Lambda g_{\mu\nu}=\gamma T_{\mu\nu} \ \ , \ \ \mu,\nu=t,x,\varphi,z
\end{equation}
read:
\begin{eqnarray}
\label{N1}
\frac{(LNN')'}{N^2 L}&=&-{\bar\Lambda}+\gamma\left[\frac{n^2(P'(x))^2}
{2\alpha L^2}-\frac14(h^2(x)-1)^2-\frac{\beta_2^2}4(f^2(x)-q^2)^2\right.
\nonumber\\
&-&\left.\frac{\beta_3^2}2
(h^2(x)-1)(f^2(x)-q^2)\right]
\end{eqnarray}
and
\begin{eqnarray}
\label{N2}
\frac{(N^2L')'}{N^2L}&=&-{\bar\Lambda}-\gamma\left[\frac{n^2h^2(x)P^2(x)}
{L^2(x)}+\frac{m^2f^2(x)}{L^2(x)}+\frac{n^2(P'(x))^2}{2\alpha L^2(x)}+
\frac14(h^2(x)-1)^2
\right.\nonumber\\
&+&\left.\frac{\beta_2^2}4(f^2(x)-q^2)^2+\frac{\beta_3^2}2
(h^2(x)-1)(f^2(x)-q^2)\right]
\end{eqnarray}
Moreover, defining $u=\sqrt{-g}=N^2L$ we get the following equation:
\begin{eqnarray}
\label{ueq}
\frac{u''(x)}{u(x)}&=&-3{\bar\Lambda}-\gamma\left[\frac{n^2h^2(x)P^2(x)}
{L^2(x)}+\frac{m^2f^2(x)}{L^2(x)}-\frac{n^2(P'(x))^2}{2\alpha L^2(x)}\right.
\nonumber\\
&+&\frac34(h^2(x)-1)^2+\frac{3\beta_2^2}4(f^2(x)-q^2)^2\nonumber\\
&+&\left.\frac{3\beta_3^2}2(h^2(x)-1)(f^2(x)-q^2)\right] \ .
\end{eqnarray}
The prime now denotes the derivatives with respect to $x$.

\subsection{Boundary Conditions}

The requirement of regularity at the origin leads to the  following boundary 
conditions:
\begin{equation}
h(0)=0, \ f(0)=0 \ , \ P(0)=n \ 
\end{equation}
for the matter fields and 
\begin{equation}
\label{zero}
N(0)=1, \ N'(0)=0, \ L(0)=0 \ , \ L'(0)=1 \ .
\end{equation}
for the metric fields. Since a cosmological horizon appears naturally in de Sitter 
space, we integrate the equations only up to this value of the coordinate $x$,
$x=x_0$. In order for the core of the local string to lie
completely within the horizon we require:
\begin{equation}
h(x=x_0)=1, \ f(x=x_0)=q \ , \ P(x=x_0)=0 \ .
\end{equation}
Note that due to the fact that the $x$ interval is finite, this is not (like in asymptotically
flat space) a necessary condition for finite energy solutions. However, we have chosen
these boundary conditions such that the energy-momentum tensor vanishes at $x=x_0$.
In addition, the limit $\bar{\Lambda}\rightarrow 0$ which leads to $x_0\rightarrow \infty$ 
can be taken with these boundary conditions. 

\section{Nielsen-Olesen strings in de Sitter space}

First, we are interested in the case of the pure Nielsen-Olesen string.
This corresponds to setting $f(x)\equiv 0$ and $q\equiv 0$ in the previous equations.
\subsection{Vacuum solution}

For the case of the pure gauge string, there is a vacuum solution of the equations.
Setting $P(x)=0$ and $h(x)=1$, we find from (\ref{ueq}), that:
\begin{equation}
N^2(x)L(x)=A\sin(\sqrt{3\bar\Lambda}x)+B\cos(\sqrt{3\bar\Lambda}x) \ , \ \ A, B \ \ {\rm constants} \ .
\end{equation}
Using the boundary conditions (\ref{zero}), we find the following solution:
\begin{equation}
N^2(x)L(x)=\frac{1}{\sqrt{3\bar\Lambda}}\sin(\sqrt{3\bar\Lambda}x) \ .
\end{equation}
This then can be put into (\ref{N1}) and (\ref{N2}) and we find the solutions:
\begin{equation}
N(x)=\cos^{2/3}(\sqrt{3\bar\Lambda}\frac x2)
\label{NON}
\end{equation}
and
\begin{equation}
L(x)=\frac{2^{2/3}}{\sqrt{3\bar\Lambda}}
[\sin(\sqrt{3\bar\Lambda} x)]^{1/3}[\tan(\sqrt{3\bar\Lambda}
\frac{x}{2})]^{2/3}
\label{NOL}
\end{equation}
where the coefficients again result from the
boundary conditions (\ref{zero}). The first zero of $N(x)$ lies at
$x^v_0=\pi/\sqrt{3\bar\Lambda}$. At the same time, 
$L(x\rightarrow x^v_0)\rightarrow \infty$. This is the cosmological horizon
of the vacuum solution. If we expand the metric functions around this horizon, we obtain:
\begin{equation}
N(x\rightarrow x^v_0)\approx(-\frac{\sqrt{3\bar\Lambda}}{2})^{2/3} (x-x^v_0)^{2/3}+...
\end{equation}
and
\begin{equation}
L(x\rightarrow x^v_0)\approx(\frac{\sqrt{3\bar\Lambda}}{2})^{-4/3} (x-x^v_0)^{-1/3}+..  \ .
\end{equation}
This has the behaviour of a so-called Kasner solution \cite{kramer}:
\begin{equation}
ds^2=(k\rho)^{2a}dt^2-d\rho^2-C^2(k\rho)^{2(b-1)}\rho^2 d\varphi^2-(k\rho)^{2c}dz^2
\end{equation}
with $a=c=2/3$, $b=-1/3$, $k=\frac{\sqrt{3\bar\Lambda}}{2}$, $C=1$.
These type of ``closed'' solutions have been found previously \cite{laguna} for the case $\bar\Lambda=0$
and were called ``supermassive'' strings. When calculating
the Kretschmann scalar $K=R^{\mu\nu\rho\sigma}R_{\mu\nu\rho\sigma}$ one obtains \cite{laguna} that $K\propto (x-x^v_0)^{-4}$ and thus the solution has indeed a curvature singularity at $x=x^v_0$. Remarkable is that in the case of  $\bar\Lambda=0$, these type of solutions only appear for
sufficiently high vacuum expectation values (vev) of the Higgs field corresponding
to strings having formed at energy scales much higher than the 
GUT scale \cite{laguna}. For smaller values of the vev no singularity appears
and the solutions exist on the full interval $[0:\infty[$.  
Accordingly, the numerical study showed \cite{yves} that these solutions
exist for a $\gamma > \gamma_{cr}$.

\subsection{Numerical results}

Subject to the boundary conditions (\ref{zero}), we have studied the coupled 
system of equations numerically.

First, we fixed $\alpha$ and $\bar\Lambda$ to study the influence of
the gravitational coupling $\gamma$ on the solutions.
We determined the  value of $x$ at which the metric function $N(x)$ vanishes, i.e. $N(x=x_0)=0$. Our results for $\alpha=1.0$, $\bar\Lambda=0.005$ and $n=1$, $2$ are shown in Fig.~1. As expected the value of $x_0$ decreases with increasing gravitational coupling $\gamma$. Moreover, we observe a steep decrease in $x_0$ for a relatively small range of $\gamma$. We have only plotted results for $\gamma$'s corresponding to $x_0 \geq 5$ since for large $\gamma$'s the numerics
becomes increasingly difficult. The reason for this is indicated in Fig.~2, where we 
show the profiles of the metric functions $N(x)$, $L(x)$ as well as those of the matter field functions $P(x)$ and $h(x)$ for $\alpha=1.0$, $\bar\Lambda=0.005$, $n=1$ 
and two different
choices of $\gamma$. For $\gamma=1.5$, the value of $x$ at which the matter field functions reach their asymptotic values $0$ and $1$, respectively, is much smaller than the value of $x_0$. 
This means that the horizon clearly lies outside the core of the string. For $\gamma=1.7$, however, the situation is different. The value of $x$ at which $h(x)$ reaches $1$ is roughly equal to $x_0$, while $P(x)$ seems to be still greater than $0$ on the plot we present. This is due to the fact that the ``real'' solution would have a slightly higher $x_0$ at which $P(x=x_0)=0$. However, since
$L(x\rightarrow x_0)\rightarrow \infty$, it is numerically impossible to reach the final solution. Nevertheless, the plot indicates that -like $h(x)$- $P(x)$ just reaches its asymptotic value $0$ at $x=x_0$. Thus the horizon lies very close to the core of the string.

Then, we fixed $\alpha$ and $\gamma$ and determined $x_0$ in dependence on $\bar\Lambda$. Our results for $\alpha=\gamma=0.5$ and $n=1$, $2$ together with the location 
of the cosmological horizon of the vacuum solution, $x^v_0$, are given in Fig.~3. 

We clearly observe that the value of the cosmological horizon
decreases with the increase of the cosmological constant, as expected.
Moreover, an increase in the vorticity $n$ leads to a decrease of $x_0$ for
the same $\bar\Lambda$. For small $\bar\Lambda$, $x_0$ of both the 
$n=1$ and the $n=2$ solution is very close to the corresponding $x^v_0$.
This can again be explained by studying the behaviour of the
matter field functions for varying $\bar{\Lambda}$. 
As observed previously \cite{GM}, we find that for fixed $\alpha$ and $\gamma$
and increasing $\bar{\Lambda}$, the value of the coordinate $x$
at which the matter field functions reach their asymptotic values
also increases, e.g. for $\alpha=\gamma=0.5$, $n=2$ we find
that for $\bar{\Lambda}=0.001$ the value of $x$, where the gauge field function
reaches $P(x^{0.1})=0.1$ is $x^{0.1}(\bar{\Lambda}=0.001)\approx 6.2$,
while for $\bar{\Lambda}=0.005$, we find 
$x^{0.1}(\bar{\Lambda}=0.005)\approx 6.45$. This can be interpreted
as representing  a thicker string core due to an increased cosmological expansion.
Thus the cosmological horizon lies closer and closer to the core of the string
for increasing $\bar\Lambda$ and so only for small $\bar\Lambda$
the solution close to the cosmological horizon can be described by the
vacuum solution. In Fig.~4, we show the profiles of $N(x)$ and $L(x)$
for $\bar\Lambda=0.01$, $\alpha=\gamma=0.5$ and $n=2$ both for the Nielsen-Olesen string and the vacuum solution. Clearly, the solutions differ quite strongly.  

\section{Global strings in de Sitter space}

This is the case of setting $P(x)\equiv n$ and $h(x)\equiv 0$ (which implies $\eta_1\equiv 0$). The global string without a cosmological constant, i.e. $\bar\Lambda=0$ has been studied extensively \cite{cohen,harari,Gibbons,gregory1}. To have a good starting solution for the construction of dS global strings, we have reconstructed the global string  without cosmological constant. We find that the metric function $N(x)$ deviates very little from one, but that the quantity $|N(x=0)-N(x^*)|$ is increasing
with $x^*\rightarrow\infty$. Moreover, $L(x)$ grows approximately linearly with $x$.
This is the behaviour found in \cite{harari}, namely that outside the core of the global string the metric functions behave like
$N^2=1-\frac{\gamma}{2}\ln(\frac{x}{x_{gc}})$, $L^2=x^2(1-\gamma\ln(\frac{x}{x_{gc}}))$ with $x_{gc}\propto (q\beta_2)^{-1}$ being the core of the global string in our rescaled coordinates. Moreover, we recover the behaviour of the Goldstone field function $f(x\rightarrow \infty)=q-O(x^{-2})$.

\subsection{Numerical results}

For non-vanishing cosmological constant we find that the behaviour of  $N(x)$ and 
$L(x)$ resembles that of the metric functions in the case of the Nielsen-Olesen
string in de Sitter space (see previous section). Again, for all 
constants fixed and $\bar\Lambda$ varied, we find that
the value at which $N(x=x_0)=0$ decreases with increasing
$\bar\Lambda$, e.g. for $\beta_2=1$, $q=0.1$, $\gamma=0.1$, we find that
$x_0(\bar\Lambda=10^{-4})\approx 180$, while $x_0(\bar\Lambda=10^{-3})\approx 57$.
At the same time, the value of $x$ at which the function $f(x)$ reaches its
vev increases with increasing $\bar\Lambda$ which is due to 
the increased cosmological expansion thus leading to an extended string core.
The behaviour of the function $f(x)$ depends crucially on the
cosmological constant which determines the cosmological horizon
and the parameters $\beta_2$ and $q$, which determine the radius of the string core.
We will discuss these features in more detail in the context of composite topological
defects in the next section.

\section{Composite system of global and Nielsen-Olesen string}
Unlike in the case of the ``pure'' Nielsen-Olesen string, a complete analytical analysis 
of the composite system seems to be impossible.
However, some additional information can  be gained by analysing the energy density per unit 
length. Before discussing our numerical results, we make some remarks on this point in the
following section.

The energy density per unit length of the composite defect is given by \cite{Christensen,yves}:
\begin{equation}
{\cal E}=\int \sqrt{-g_3} T^0_0 dx_1 dx_2
\end{equation}
where $g_3$ is the determinant of the $2+1$ dimensional metric
$ds^2=N^2(\rho)dt^2-d\rho^2-L^2(\rho)d\varphi^2$ and $T^0_0=-{\cal L}_M$ is the 
$00$-component of the energy-momentum tensor. In the cylindrical coordinates,
we get:
\begin{eqnarray}
\label{E}
{\cal E}&=&\pi\eta_1^2\int_0^{x_0}d x N(x)L(x)\left[
\frac{n^2}{\alpha L^2(x)}(P'(x))^2+(h'(x))^2+(f'(x))^2\right.\nonumber\\
&+&\frac{n^2h^2(x)P^2(x)}{L^2(x)}+\frac{m^2f^2(x)}{L^2(x)}+
\frac12(h^2(x)-1)^2+\frac{\beta_2^2}2(f^2(x)-q^2)^2\nonumber\\
&+&\left.\beta_3^2(h(x)^2-1)(f^2(x)-q^2)\right] \ .
\end{eqnarray}
From this, we can observe that the composite defect 
has a finite energy density. There are two different scenarios now. If the product 
$\beta_2 q$ is large enough the core of the global string is small and
the cores of both strings lie within the cosmological horizon $x_0$. We can then 
assume that $f(x_0)=q$, $f^{'}(x_0)=0$, $P(x_0)=0$, $h(x_0)=1$.
From before, we know that $L(x)$ goes to infinity close to the horizon, while
$N(x)L(x)$ remains finite. Thus the integrand of (\ref{E}) tends to zero like
$m^2q^2\frac{N(x)}{L(x)}$. If the product $\beta_2 q$ is small, the
core of the global string extends to outside the horizon $x_0$. 
Then the integrand of (\ref{E}) becomes 
$N(x)L(x)\left[(f'(x))^2+\frac{m^2f^2(x)}{L^2(x)}+\frac{\beta_2^2}2(f^2(x)-q^2)^2\right]$. 
Since close to the horizon, the product $N(x)L(x)$ tends to
zero, this is finite. We have indeed confirmed numerically that this is the case.

\subsection{Numerical Results}

Assuming that the behaviour of the metric functions $N(x)$ and $L(x)$ persists in the presence 
of all (non-trivial) matter fields (which indeed our numerical analysis confirms),
we can insert (\ref{NON}) and (\ref{NOL}) into (\ref{eqf}). We obtain that:
\begin{equation}
f(x\rightarrow x_0)\sim q+ C(x-x_0)^{8/3} \ \ , \ \ \ C \ \ {\rm constant}
\end{equation}

We have construct the composite model solution
by starting from the Nielsen-Olesen (NO) string and increasing the parameter
$q$ gradually from zero.
Our numerical analysis confirms the assumption that the radius of the NO
string core is smaller than that of the global string. Moreover, we find
that the NO string always resides inside 
the horizon.
As for the global string, we find that when the coupling constant $\beta_2$ is small
the function $f(x)$ 
reaches the imposed expectation value  $f(x_0) = q$ with a positive
concavity, in particular the derivative $f'|_{x=x_0}$ is non zero
and our numerical solution clearly suggests that the behaviour
$f(x\rightarrow x_0) \sim q + C (x - x_0)^{8/3}$ is not fulfilled.
This is demonstrated in Fig.~5, where we show $f(x)$ of the composite 
defect for $\beta_2^2=1$, $\alpha=\gamma=\beta_1^2=1$, $q=0.1$, $\beta_3^2=0$ and
$\bar\Lambda=0.001$. 
This suggests that for small values of the product $\beta_2 q$, the argument demonstrated
above doesn't hold.

Increasing the value of $\beta_2$ we were able to 
produce solutions which seem to have the expected behaviour, i.e. 
$f(0) =0$, $f(x=x_0)=q$ and $f'|_{x=x_0}=0$. This is shown in Fig.~5
for $\alpha=\gamma=\beta_1^2=1$, $q=0.1$, $\beta_3^2=0$,
$\bar\Lambda=0.001$ and $\beta_2^2=5$, $10$, $20$, $50$, respectively.
Of course, the occurrence
of the singularity at $x = x_0$ renders the interpretation of the numerical results 
not hundred percent certain but we are rather confident that
composite string defects which are regular inside the horizon
exist for large enough $\beta_2$.

All these results are obtained for the case of $\beta_3=0$, i.e. the two defects
interact with each other only indirectly over gravity.
We have also attempted to construct solutions with $\beta_3\neq 0$.
We find that only for large enough values of the quotient $\beta_2^2/\beta_3^2$,
the solutions seem well behaved. If $\beta_2^2/\beta_3^2$ is roughly of the order
of $10^2$ (for $\alpha=\gamma=\beta_1^2=1$, $q=0.1$ and $\bar\Lambda=0.001$),
the behaviour of the functions is 
very similar to that in the case of $\beta_3=0$. For $\beta_2^2/\beta_3^2$
smaller than that, however, the functions $f(x)$ and $h(x)$ start to develop
oscillations close to the cosmological horizon. The number of oscillations
increases with the decrease of the quotient $\beta_2^2/\beta_3^2$.
Thus, we conclude that directly interacting 
composite defects without the global string singularity only 
exist if the self-interaction of the global string
is much larger than the interaction between the two defects.

\section{Conclusion}

In this paper we have analysed both Nielsen-Olesen and global strings as
well as the composite system of both defects in de Sitter space. 
When the matter fields are set equal to their vacuum expectation
values (vev) in the case of the ``pure'' Nielsen-Olesen string, 
we were able to construct analytic solutions of the Einstein equations
in terms of trigonometric functions. The metric tensor resembles that
of so-called ``supermassive'' strings which exist in asymptotically
flat space only for sufficiently high enough values of the
vev of the Higgs field \cite{laguna}. These were considered as being ``unphysical''
since they should have formed at energy scales high above the GUT scale. Since recent
observations indicate that we live in a universe with positive cosmological constant,
and since we find that the existence of our solutions is not restricted to
values of the Higgs field's vev being large enough, these solutions might well
be of relevance.

Our numerical analysis suggests that the general behaviour of the vacuum metric
persists in the presence of the matter fields. E.g.  comparing the location of the
cosmological horizon $x_0$ in dependence on the coupling constants for the ``pure'' Nielsen-Olesen
string  and that of the vacuum solution, $x^v_0$, we find that for small $\bar\Lambda$ and/or $\gamma$,
$x_0$ and $x^v_0$ are nearly equal. For increasing $\bar\Lambda$ and/or $\gamma$, the difference
between the two increases. The reason for this is that the radius of the string core
becomes comparable to the radius of the cosmological horizon and thus the assumption of a ``vacuum''
at the cosmological horizon is not valid any longer.

Constructing the global string and the composite defect of Nielsen-Olesen and global string,
we find that the existence of  solutions without a singularity resulting from the
global string itself depends crucially
on the product $\beta_2 q$ which (in our rescaled coordinates) is inverse proportional to
the radius of the global string's core. If (for fixed $q$) $\beta_2$ is too small, the core
of the global string extends to outside the cosmological horizon and the
function $f$ reaches its vev with positive concavity. For large enough values of $\beta_2$,
the core of the global string lies inside the horizon and
our numerical results seem to indicate that a global string/composite
defect without the normal singularity of the global string exists. 
However, note that
the removal of the global string
singularity which exists in asymptotically flat space
can be achieved only by introducing a ``new'' singularity, the curvature singularity
at the horizon.
\\
\\
\\
\\
\\
{\bf Acknowledgement} YB acknowledges the Belgian FNRS
for financial support. BH was supported by an EPSRC grant.\\
\\
\\
{\bf Note added} After finishing the manuscript, B. Linet has brought to our intention
his paper ``The static, cylindrically symmetric strings in general relativity with cosmological
constant'' [J. Math. Phys. {\bf 27} (1986), pp. 1817-1818], in which he discusses the vacuum
solutions (\ref{NON}) and (\ref{NOL}) ``re-found'' by us.

\newpage

\begin{figure}

\centering

\epsfysize=10cm

\mbox{\epsffile{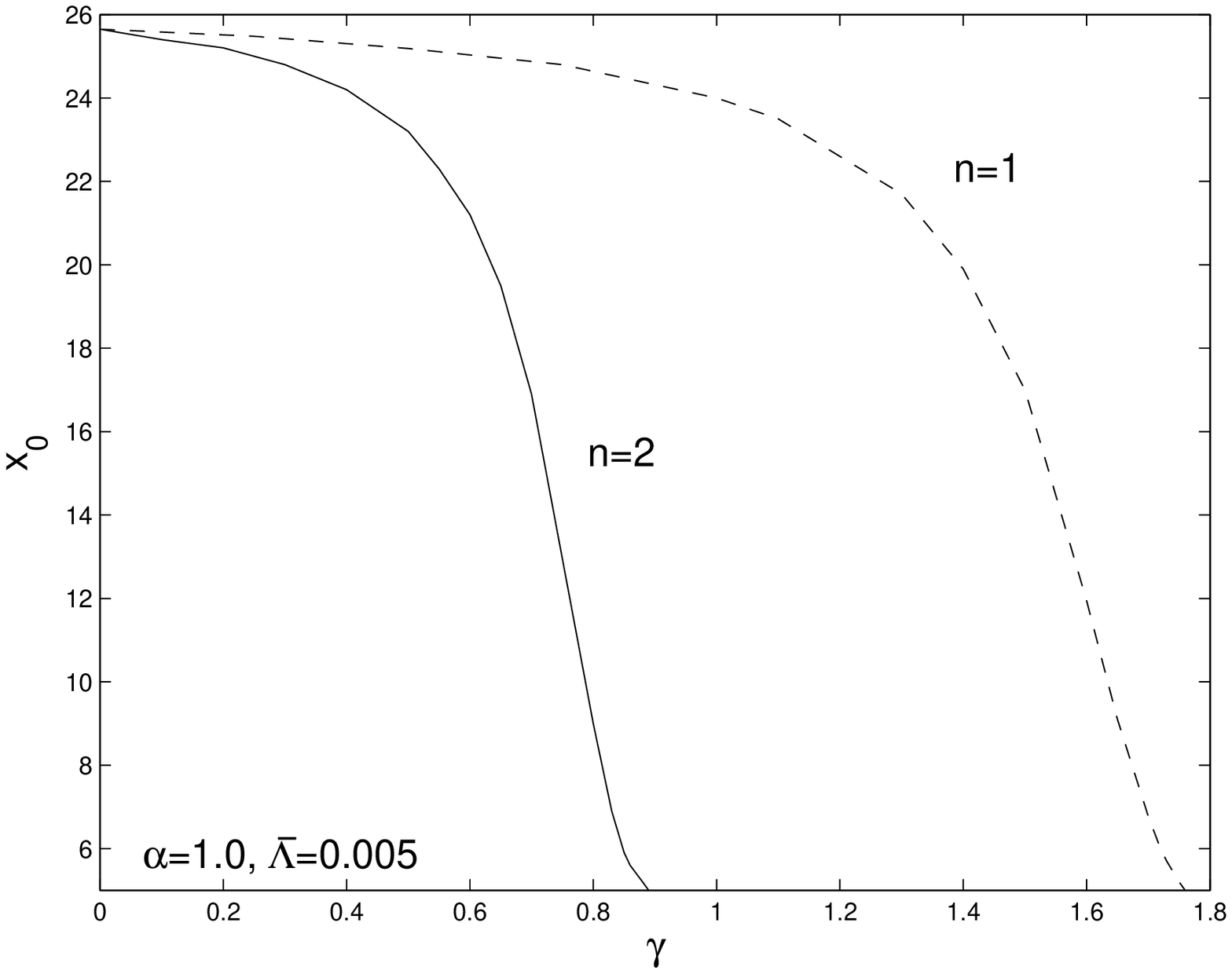}}

\caption{The value of the dimensionless coordinate $x$ at which a cosmological
horizon appears, $x=x_0$, is given for the Nielsen-Olesen string as function 
of $\gamma$ for $n=1$ (dashed) and $n=2$ (solid) with $\alpha=1.0$ and $\bar\Lambda=0.005$. }

\end{figure}

\newpage

\begin{figure}

\centering

\epsfysize=10cm

\mbox{\epsffile{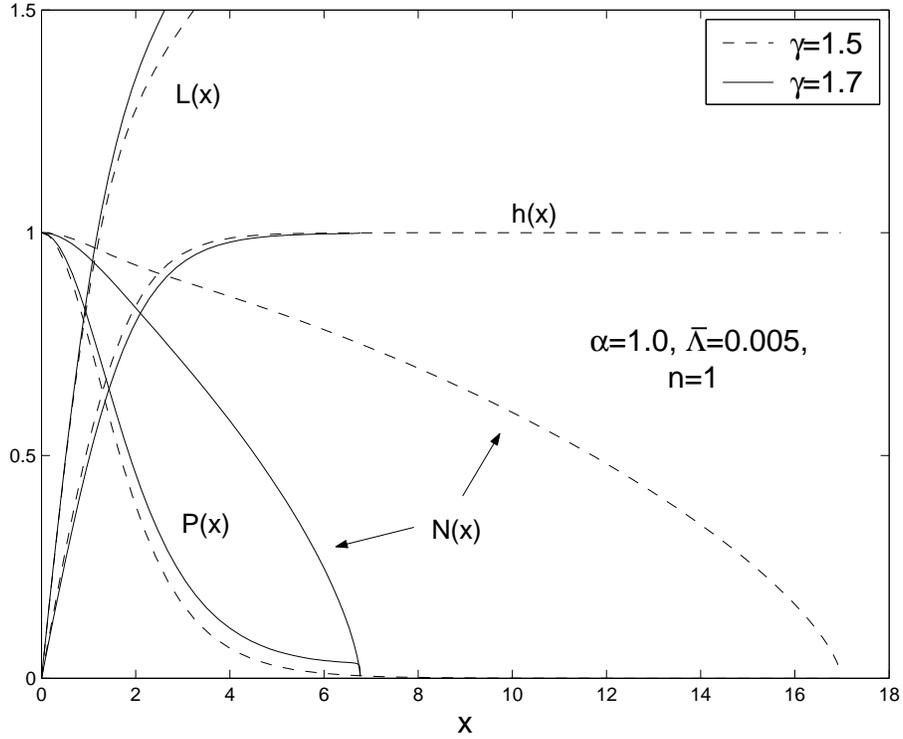}}

\caption{The profiles of the metric functions $N(x)$, $L(x)$ and the profiles of
the matter field functions $P(x)$ and $h(x)$ are shown for $\alpha=1.0$,
$\bar\Lambda=0.005$, $n=1$ and $\gamma=1.5$ (dashed) and $\gamma=1.7$ (solid), respectively. }

\end{figure}

\newpage

\begin{figure}

\centering

\epsfysize=10cm

\mbox{\epsffile{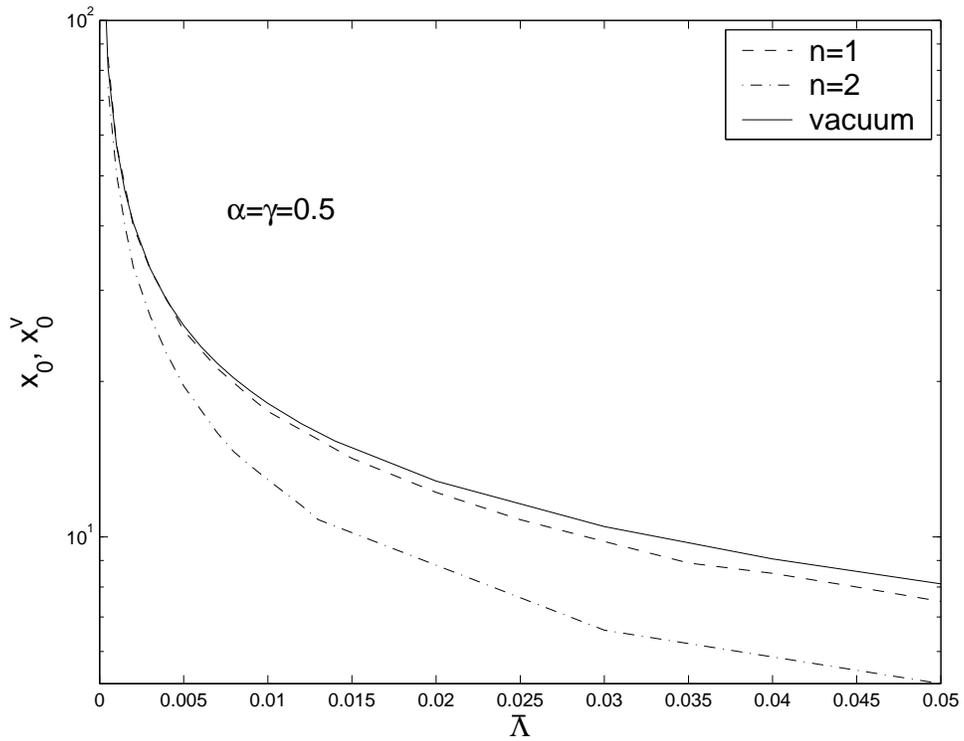}}

\caption{The value of the dimensionless coordinate $x$ at which a cosmological
horizon appears, $x=x_0$, is given for the Nielsen-Olesen string
as function  of $\bar\Lambda$ for 
$n=1$ and $n=2$ with $\alpha=\gamma=0.5$. For comparison, also the value 
$x^v_0=\pi/\sqrt{3\bar\Lambda}$ for the vacuum solution is given. }

\end{figure}

\newpage

\begin{figure}

\centering

\epsfysize=10cm

\mbox{\epsffile{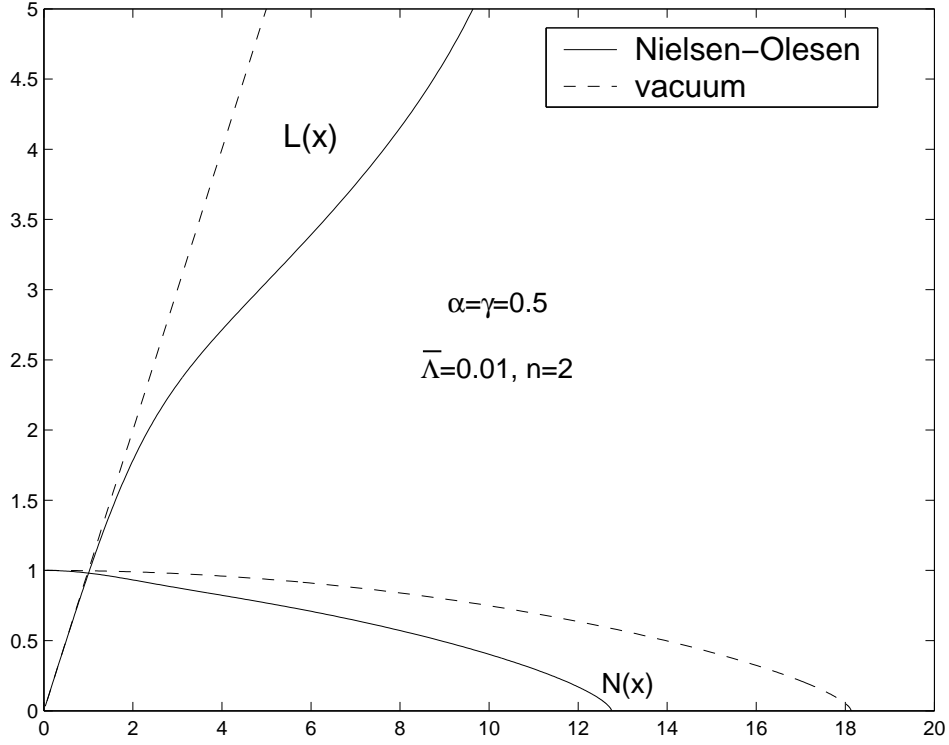}}

\caption{The profiles of the metric functions $N(x)$ and $L(x)$ 
 are shown for $\alpha=\gamma=0.5$,
$\bar\Lambda=0.01$ and vorticity $n=2$. We compare the Nielsen-Olesen solution
(solid) with the corresponding vacuum solution (dashed).}
\end{figure}

\newpage
\begin{figure}
\centering
\epsfysize=10cm
\mbox{\epsffile{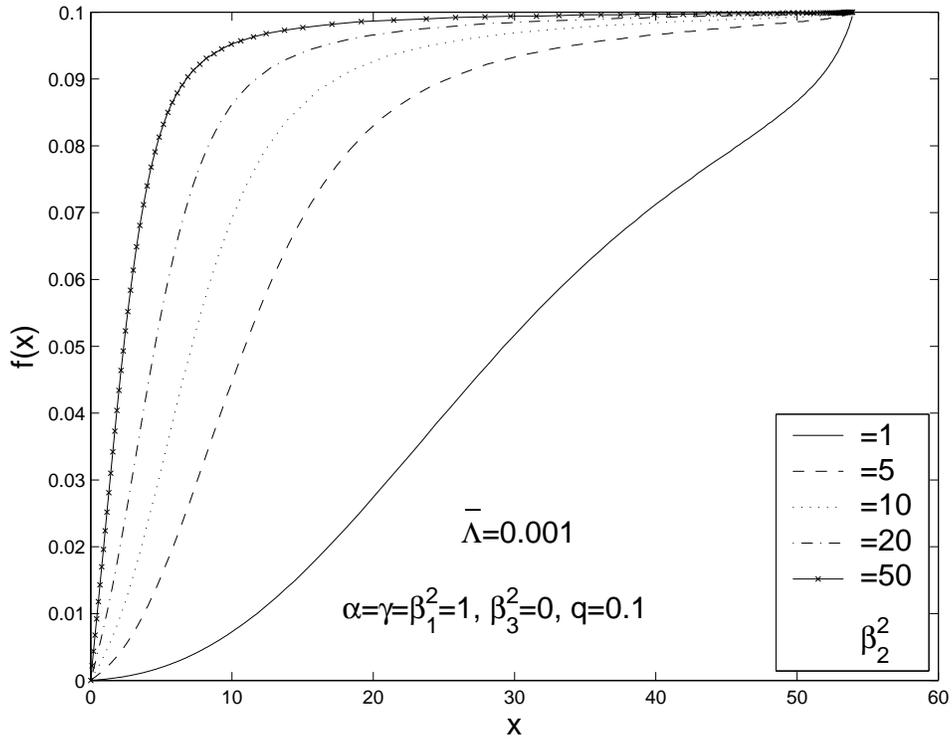}}
\caption{The profile of the Goldstone field function $f(x)$  
is shown for the composite defect with $\alpha=\gamma=\beta_1^2=1$, $\beta_3^2=0$, $\bar\Lambda=0.001$, $q=0.1$,
vorticity $n=1$ and different values of $\beta_2^2$.}
\end{figure}

\end{document}